# Ferromagnetism in the Highly-Correlated Hubbard Model

Valentin Yu. Irkhin[1, a] and Alexander V. Zarubin[1, b]

[1]Russia, Yekaterinburg, Institute of Metal Physics, S. Kovalevskaya str. 18

[a]Valentin.Irkhin@imp.uran.ru, [b]Alexander.Zarubin@imp.uran.ru

**Abstract.** The Hubbard model with strong correlations is treated in the many-electron representation of Hubbard's operators. The regions of stability of saturated and non-saturated ferromagnetism in the *n*–*U* plane for the square and simple cubic lattices are calculated. The role of the bare density of states singularities for the magnetic phase diagram is discussed. A comparison with the results of previous works is performed.

**Introduction**

The problem of itinerant-electron ferromagnetism in narrow bands is important for a number of highly-correlated *d*-electron systems (see [1,2]), especially for half-metallic ferromagnets [3] which (in the simple one-band Hubbard model [4]) are just saturated ferromagnets.

As first demonstrated by Nagaoka [5], in the limit of infinite Hubbard's repulsion $U$ the ground state for simple lattices is a strong (saturated) ferromagnetic state for a low density δ of current carriers (doubles or holes in an almost half-filled band). Further on, Nagaoka found the spin-wave instability of the saturated ferromagnetism with increasing δ and decreasing $U$. Roth [6] applied a variational principle to this problem and obtained two critical concentrations. The first one, $\delta_c$, corresponds to instability of saturated ferromagnetic state, and the second one, $\delta_c'$, to the transition from non-saturated ferromagnetism into paramagnetic state.

Next, the region of stability of the saturated ferromagnetic was investigated within various approximations in numerous works (see [8–11] and a discussion in Refs. [3,7]). In particular, a variational method has been used [9] to provide a rigorous boundary of saturated ferromagnetism for a square lattice. On the other hand, Hanish and Igarashi [10] obtained the boundary of instability of the paramagnetic state for the same lattice.

It is difficult to consider reliably the limit of strong correlations in the standard Stoner-like one-electron approaches since they do not correctly treat the formation of local moments [1]. However, simple approximations like "Hubbard-I" [4] do not yield a correct description of ferromagnetism because of violation of kinematical requirements [7,2]. In our previous papers [7] we have applied the $1/z$-expansion of the Green's functions in the many-electron representation for the $U = \infty$ Hubbard model and obtained a rather simple interpolation description of saturated and non-saturated ferromagnetism. In the present work we generalize this consideration to the finite-$U$ case to obtain a more full phase diagram.

**Calculation of the One-Particle Green's Functions in the Hubbard Model**

We start from the Hamiltonian for the Hubbard model in the many-electron $X$-operator representation [12,13]

$$H = \sum_{\mathbf{k}\sigma} t_{\mathbf{k}} (X_{\mathbf{k}}^{\sigma 0} + \sigma X_{\mathbf{k}}^{2-\sigma})(X_{-\mathbf{k}}^{0\sigma} + \sigma X_{-\mathbf{k}}^{-\sigma 2}) + U \sum_i X_i^{22}, \qquad (1)$$

where $t_\mathbf{k}$ is the band energy, $X_\mathbf{k}^{\alpha\beta}$ are the Fourier transforms of Hubbard's operators,

$$X_i^{\alpha\beta} = |i\alpha\rangle\langle i\beta|, \qquad X_i^{\alpha\beta} X_i^{\gamma\varepsilon} = \delta_{\beta\gamma} X_i^{\alpha\varepsilon}, \tag{2}$$

($\alpha = 0$ denotes holes, $\alpha = 2$ doubles, and $\alpha = \sigma = \pm(\uparrow,\downarrow)$ singly occupied states on a site). They are expressed in terms of one-electron Fermi operators as

$$X_i^{\sigma 0} = c_{i\sigma}^\dagger (1 - n_{i-\sigma}), \qquad X_i^{2-\sigma} = \sigma c_{i\sigma}^\dagger n_{i-\sigma}. \tag{3}$$

We calculate the one-electron Green's function

$$G_{\mathbf{k}\sigma}(E) = \langle\langle c_{\mathbf{k}\sigma} | c_{\mathbf{k}\sigma}^\dagger \rangle\rangle_E. \tag{4}$$

In the Hubbard-I approximation (zeroth order in the inverse nearest-neighbor number $1/z$) we have in the locator representation

$$G_{\mathbf{k}\sigma}^0(E) = [F_\sigma^0(E) - t_\mathbf{k}]^{-1}, \qquad F_\sigma^0(E) = E(E-U)/(E - U(n_0 + n_\sigma)), \qquad n_\alpha = \langle X_i^{\alpha\alpha}\rangle. \tag{5}$$

Decoupling the sequence of equations of motion similar to Ref. [7] we derive with account of spin and charge fluctuations the self-consistent equations

$$G_{\mathbf{k}\sigma}(E) = [F_{\mathbf{k}\sigma}(E) - t_\mathbf{k}]^{-1}, \qquad F_{\mathbf{k}\sigma}(E) = b_{\mathbf{k}\sigma}(E)/a_{\mathbf{k}\sigma}(E), \tag{6}$$

$$a_{\mathbf{k}\sigma}(E) = [F_\sigma^0(E)]^{-1} + \left(\frac{1}{E} - \frac{1}{E-U}\right)\sum_\mathbf{q} t_\mathbf{q}\left(\frac{-En_{\mathbf{q}-\sigma} + U(\langle c_{\mathbf{q}-\sigma}^\dagger X_\mathbf{q}^{0-\sigma}\rangle - \chi_{-\mathbf{k}+\mathbf{q}}^{-\sigma\sigma})}{E - U(n_0 + n_{-\sigma})} G_{\mathbf{q}-\sigma}(E) + \right.$$
$$\left. + \frac{En_{\mathbf{q}-\sigma} + U(-\sigma\langle X_{-\mathbf{q}}^{2\sigma} c_{\mathbf{q}-\sigma}\rangle - \kappa_{-\mathbf{k}+\mathbf{q}}^{-\sigma\sigma})}{E - U(n_\sigma + n_2)} \widetilde{G}_{\mathbf{q}\sigma}(E)\right) - U\left(\frac{1}{E} - \frac{1}{E-U}\right)\sum_\mathbf{q} t_\mathbf{q} \frac{\chi_{-\mathbf{k}+\mathbf{q}}^{zz}}{E - U(n_0 + n_{-\sigma})} G_{\mathbf{q}\sigma}(E),$$

$$b_{\mathbf{k}\sigma}(E) = 1 - \left(\frac{1}{E} - \frac{1}{E-U}\right)\sum_\mathbf{q} t_\mathbf{q}^2 \left(\frac{En_{\mathbf{q}-\sigma} + U\langle c_{\mathbf{q}-\sigma}^\dagger X_\mathbf{q}^{0-\sigma}\rangle}{E - U(n_0 + n_{-\sigma})} G_{\mathbf{q}-\sigma}(E) + \frac{En_{\mathbf{q}-\sigma} + \sigma U\langle X_{-\mathbf{q}}^{2\sigma} c_{\mathbf{q}-\sigma}\rangle}{E - U(n_\sigma + n_2)} \widetilde{G}_{\mathbf{q}\sigma}(E)\right),$$

where the occupation numbers $n_{\mathbf{q}\sigma} = \langle c_{\mathbf{q}\sigma}^\dagger c_{\mathbf{q}\sigma}\rangle$ are determined self-consistently,

$$\chi_\mathbf{q}^{\sigma-\sigma} = \langle S_\mathbf{q}^\sigma S_{-\mathbf{q}}^{-\sigma}\rangle, \quad \kappa_\mathbf{q}^{\sigma-\sigma} = \langle \rho_\mathbf{q}^\sigma \rho_{-\mathbf{q}}^{-\sigma}\rangle, \quad \chi_\mathbf{q}^{zz} = \langle \delta(\sigma S_\mathbf{q}^z - \rho_\mathbf{q}^z)\delta(\sigma S_{-\mathbf{q}}^z - \rho_{-\mathbf{q}}^z)\rangle \tag{7}$$

are spin and charge correlation functions,

$$S_\mathbf{q}^\sigma = X_\mathbf{q}^{\sigma-\sigma}, \quad S_\mathbf{q}^z = \tfrac{\sigma}{2}(X_\mathbf{q}^{\sigma\sigma} - X_\mathbf{q}^{-\sigma-\sigma}), \quad \rho_\mathbf{q}^\sigma = X_\mathbf{q}^{20}, \quad \rho_\mathbf{q}^z = \tfrac{1}{2}(X_\mathbf{q}^{22} - X_\mathbf{q}^{00}). \tag{8}$$

The expression for $\widetilde{G}_{\mathbf{q}\sigma}(E)$ reads

$$\widetilde{G}_{\mathbf{k}\sigma}(E) = [\widetilde{F}_{\mathbf{k}\sigma}^0(E) + t_\mathbf{k}]^{-1}, \qquad \widetilde{F}_{\mathbf{k}\sigma}(E) = b_{\mathbf{k}\sigma}(E)/\widetilde{a}_{\mathbf{k}\sigma}(E), \tag{9}$$

where $\widetilde{a}_{\mathbf{k}\sigma}(E)$ differs from $a_{\mathbf{k}\sigma}(E)$ by the replacement

$$F_\sigma^0(E) \to \widetilde{F}_\sigma^0(E) = E(E-U)/(E - U(n_\sigma + n_2)). \tag{10}$$

Unlike the local Hubbard-III approximation [14], the expressions (6)–(10) enable one to treat consistently the effects of Fermi excitations.

**Magnetic Phase Diagram**

The ferromagnetic state in the simple Hubbard model occurs for large $U$ only, so that we can neglect charge fluctuations and put $N_2 = 0$ for the electron concentration $n < 1$ (we have in the large-$U$ limit $N_2 \propto 1/U^2$). This enables us to fix the number of holes $N_0 = \delta = 1 - n$ and solve the self-consistency equations for the ground-state magnetization $m = \langle S^z \rangle$ and the chemical potential only.

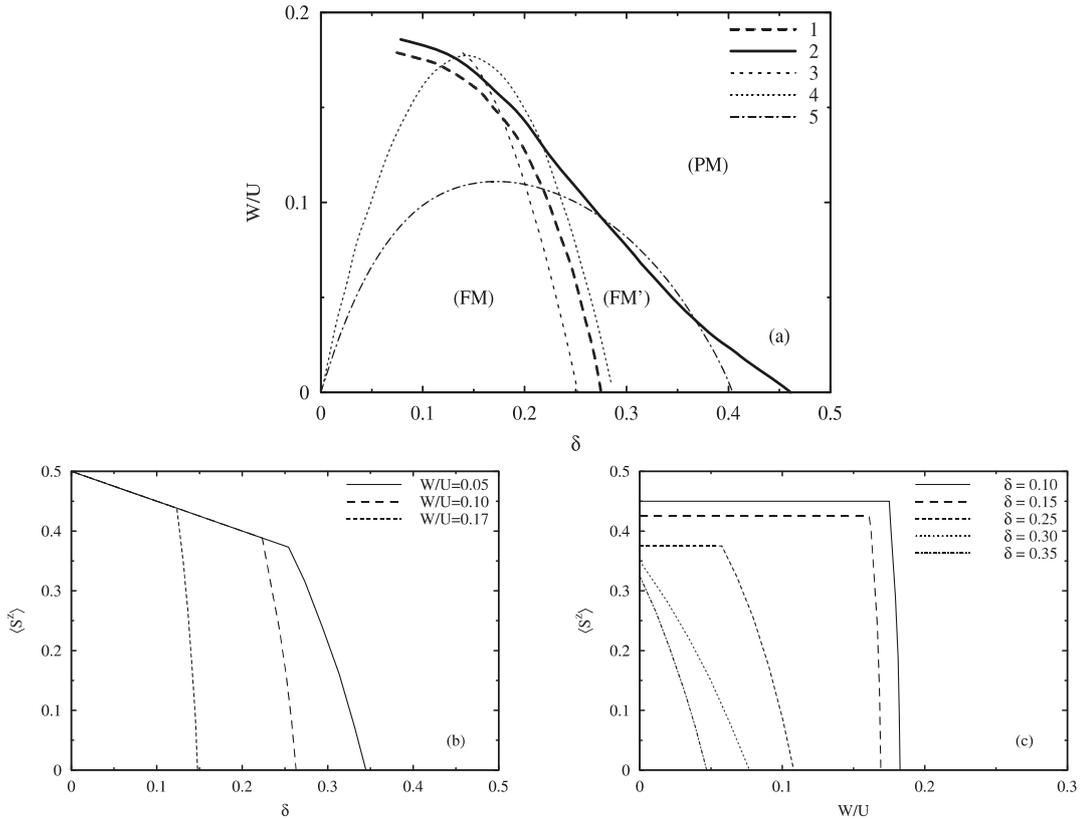

Fig. 1. Magnetic phase diagram for the square lattice (a) (*line 1* is saturated to non-saturated ferromagnetic transition, *line 2* is ferromagnetic to paramagnetic transition, *line 3* is the Edwards–Hertz approximation [8], *line 4* is the von der Linden–Edwards result [9], and *line 5* is the Hanisch–Igarashi result [10]), and the corresponding dependences of the ground-state magnetization vs. hole concentration (b) and inverse Coulomb interaction $W/U$ ($W$ is half-bandwidth) (c)

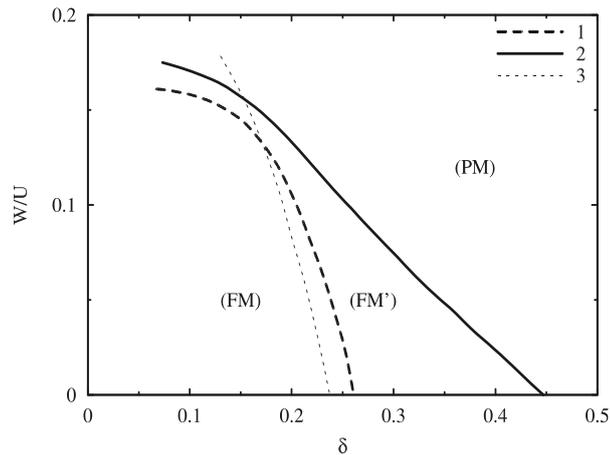

Fig. 2. Magnetic phase diagram for the simple cubic lattice (*line 1* is saturated to non-saturated ferromagnetic transition, *line 2* is ferromagnetic to paramagnetic transition, and *line 3* is the Edwards–Hertz approximation [9])

The results of the numerical solution of Eqs. (6)–(10) are shown in Figs. 1–2. The phase diagram demonstrates two magnetic phase transitions: from saturated ferromagnetism (FM) to non-saturated ferromagnetism (FM'), and from non-saturated ferromagnetism to paramagnetic state (PM).

The results for the first transition (FM–FM') are compared with results of previous works yielding a satisfactory agreement.

One can see that for the square lattice the region of non-saturated ferromagnetism decreases rapidly with decreasing $U$, so that the ground-state magnetic transition is practically of first order. This is due to the strong singularity of the bare density of states in the middle of the band. Since two peaks with opposite spin projections tend to be separated, the non-saturated ferromagnetism is energetically non-favorable. This fact ensures also the small second critical concentration in the square lattice in comparison with the other lattices. When introducing the next-nearest-neighbor electron transfer $t'$, the singularity is shifted to the band bottom, so that the critical values of $U$ decrease, as well as in the mean-field approximation [15].

To conclude, we have obtained the magnetic phase diagram of the large-$U$ Hubbard model, which is in agreement with previous approaches. The investigations of more realistic three-dimensional lattices (in particular, with the density-of-states Van Hove singularities) within our approach would be of interest. A consistent consideration should also include antiferromagnetic and phase-separated states which occur at smaller $U$ and $\delta$.

The work is supported in part by Presidium of RAS Program "Quantum Physics of Condensed Matter".